\documentclass[twocolumn,prb]{revtex4}
\usepackage{graphicx}

\newcommand{\beq}{\begin{equation}}
\newcommand{\eeq}{\end{equation}}
\newcommand{\beqa}{\begin{eqnarray}}
\newcommand{\eeqa}{\end{eqnarray}}

\newcommand{\qav}[1]{\left\langle #1 \right\rangle}
 \newcommand{\rem}[1]{}
\newcommand{\refe}[1]{(\ref{#1})}

\newcommand{\refE}[1]{Eq.~(\ref{#1})}
 \newcommand{\Pc}{{\cal P}}
 \newcommand{\larghezza}{6.cm}
 \newcommand{\Qc}{{\cal Q}}
\newcommand{\Ac}{{\cal A}} \newcommand{\Dc}{{\cal D}}
\newcommand{\Fc}{{\cal F}}

\begin{document}

\title{Self-consistent theory of molecular switching}

\author{F. Pistolesi}
\affiliation{Universit\'e Joseph Fourier,\\
Laboratoire de Physique et Mod\'elisation des Milieux Condens\'es,\\
C.N.R.S. B.P. 166, 38042 Grenoble, France}

\author{Ya. M. Blanter}
\affiliation{Kavli Institute of Nanoscience, Delft University of Technology, Lorentzweg 1, 2628\\
CJ Delft, The Netherlands}

\author{Ivar Martin}
\affiliation{Theoretical Division, Los Alamos National Laboratory, \\
Los Alamos, New Mexico, 87544, USA}

\date{\today}
\begin{abstract}
We study the model of a molecular switch comprised of a molecule with a soft
vibrational degree of freedom coupled to metallic leads.  In the presence of
strong electron-ion interaction, different charge states of the molecule
correspond to substantially different ionic configurations, which can lead to
very slow switching between energetically close configurations (Franck-Condon
blockade).  Application of transport voltage, however, can drive the molecule
far out of thermal equilibrium and thus dramatically accelerate the switching.
The tunneling electrons play the role of a heat bath with an effective
temperature dependent on the applied transport voltage.   Including the
transport-induced ``heating" selfconsistently, we determine the stationary
current-voltage characteristics of the device, and the switching dynamics for
symmetric and asymmetric devices. We also study the effects of an extra
dissipative environment and demonstrate that it can lead to enhanced
non-linearities in the transport properties of the device and dramatically
suppress the switching dynamics.

\end{abstract}
\maketitle

\section{introduction}
\label{Sec:intro}

The apparent limitations of the silicon-based technology on the way to further
acceleration and miniaturization have prompted active research into alternative
electronic architectures.\cite{molecular,hybrid_Likharev}  In particular,
molecular electronics holds a lot of promise because each molecule, being only
about a nanometer in size can in principle perform such non-trivial operations
as information storage\cite{mem} or electrical current
rectification.\cite{mol_diod} Since molecules, even intricate ones, can be mass
produced by means of well-controlled chemical synthesis, one expects them to be
less susceptible to the issues of disorder that plague the silicon-based
electronics below the 10 nm scale.  The ultra-miniaturization that molecular
electronics affords, however, also leads to the problem of connecting the
molecular elements among each other, as well as of the necessary interfacing
with large-scale conventional electronics.  Indeed, early on, this problem has
caused many difficulties in reproducing results from one device to
another.\cite{nature}  However, recent advances in
fabrication\cite{recent_expt} as well as better theoretical understanding of
physics and chemistry at the point of contact \cite{recent_th} demonstrate that
this difficulty is not fundamental and promise to make reliable and
reproducible molecular junctions a reality.

There is, however, a fundamental difference that distinguishes the molecular
devices from the conventional semiconductor ones.  For a molecule to perform
its unique function, it has to be well isolated from most environmental
influences, except for the (metallic or semiconducting) contacts that are
required to access it. Under standard operation of the device, the chemical
potentials differ by the value of the applied transport voltage $V$ mutiplied by the
electron charge $e$, and thus the
environment that the molecule experiences can not be considered as equilibrium
if the voltage is greater than the temperature, $eV > k_B T$ ($k_B$ being
Boltzmann constant).  Therefore, to
determine the behavior of a molecular device under such conditions, one needs
to determine {\em selfconsistently} the influence, e.g. of electrical current
on the molecular dynamics, and vice versa, the influence of non-thermal
vibrations or electronic excitations of the molecule on the current.  This is
very different from the conventional electronics where devices are rarely
driven out of thermal equilibrium far enough to significantly affect the
performance (exceptions are the non-linear devices, such as Gunn diod).

One of the most promising and interesting molecular devices is a switch, which
can be used for information storage.  Switching has been observed
experimentally in several molecular junctions.\cite{sw_expt}  Proposed
theoretical explanations for switching range from (a) large and small-scale
molecular conformational changes, (b) changes in the charge state of
the molecule, or (c) combination of the two,
or ``polaronic".
 The purely electronic switching mechanism (b), while
possible, appears quite impractical since it would require a separate contact
in order to change the charge state of the part of the molecule that would play
the role analogous to the floating gate in flash memory by electrostatically
affecting the ``channel" current.  The switching mechanisms (a) and (c) upon
closer inspection turn out to be fundamentally the same, since in order to be
able to switch and read out the conformational state of the molecule
electronically there necessarily has to be a coupling between the electronic
and ionic degrees of freedom.  The {\em dynamical} stability of the ``on" and
``off" states in these mechanisms is achieved due to the collective nature of
the states, which now involve not only the electronic occupancy but also all
the positions of the ions in the molecule.  Thus the change of the charge state
of the molecules is accompanied by the ionic rearrangement, which for strong
enough electron-ion coupling can dramatically slow down the charge state
switching.  This is the essence of the Franck-Condon ``blockade."
\cite{braig,vonOppen,MHM,galperin}  By chemically engineering molecules with strong
electron-ion coupling and soft (low frequency) vibrational modes one can
achieve arbitrarily slow equilibrium switching rates.

In order for molecular memory element to be useful, it has to have a long
retention time (slow switching rate in the absence of any drive) but fast write
time, i.e. it should be possible to accelerate the switching rate by gate or
transport voltages.  It is easy to see that in molecular switches with
polaronic mechanism the transport-driven switching acceleration occurs
naturally.  As soon as the transport voltage exceeds the vibrational energy
quantum, $eV > \hbar \omega_0$, additional transport, as well as switching,
channels open, which correspond to electron tunneling on and off the molecule
with simultaneous excitation of vibrational quanta.\cite{MHM,vonOppen}
Moreover, enhanced charge fluctuations on the molecule effectively ``heat up"
the molecule, further increasing the current through the device.  This leads to
a positive feed-back loop which saturates when the energy transferred to the
molecule from non-equilibrium tunneling electrons exactly balances the energy
transferred back from the molecule to electrons.  As a result, the stationary
switching rate can vastly exceed the equilibrium switching strongly suppressed
due to the Franck-Condon physics.

Most of the molecular devices studied experimentally so far have been
weakly coupled to the leads.\cite{park,yu,park2,zhit,qiu,chae,cnt}  This
corresponds to the bare tunnel broadening $\hbar \Gamma$  of molecular
electronic levels smaller that the energy required to excite one oscillator quantum (phonon) $\hbar\omega_0$.  The single-electron effects play
a crucial role in this case.  They are well theoretically described by a model of
a single-electron tunneling (SET) device coupled to a single-mode
harmonic oscillator, developed  mostly in the context of nanoelectromechanical systems.
In the strong-coupling regime, when the electron-ion interaction energy $E_p$ (defined below) exceeds  $\hbar\omega_0$, the physics is governed by the Franck-Condon
effect, {\em i.e.} when the tunneling of an electron onto the molecule with the
simultaneous emission or absorption of several phonons is more probable than
elastic tunneling.   The current as the function of voltage exhibits steps
separated by $\hbar \omega_0/e$, \cite{schoeller,braig,mitra,aji}  and
the non-equilibrium electronic heating of the molecular
vibrational mode leads to self-similar avalanche dynamics of current with the
intervals of large current alternating with the periods of strongly suppressed
current.\cite{vonOppen}

In this paper, we study the case of ``slow'' phonons at strong
coupling, $\Gamma \gg \omega_0$ for $eV>\hbar \omega_0$.\cite{galperin,MHM,armour,doiron,fabio}
The physical distinction between this case and the one of ``fast" phonons,
$\Gamma \ll \omega_0$,  can be understood in the following way.  For fast phonons,
 every electron tunneling event occurs over many oscillator periods.  Thus effectively
 electrons can only couple to (or ``measure") the energy (i.e. occupation number) of the
  oscillator.\cite{zurek,MarZur}  In the opposite regime, $\Gamma \ll \omega_0$, electron
  tunneling is fast, and thus electrons are sensitive to the position of the oscillator.
  Therefore, in the former case, as a result of electron tunneling, the oscillator density
  matrix becomes close to diagonal in occupation number basis (and thus {\em non-classical}),
   and in the latter case, it is nearly diagonal in the position basis (and thus {\em classical}).
    In Ref.   \onlinecite{MHM} it has been rigorously demonstrated {\em for arbitrary coupling}
     that the condition for the onset of the classical (Langevin) dynamics is given by $\min(\hbar\Gamma, eV) \gg \hbar\omega_0$.
Even at weak coupling, $E_p < \hbar \omega_0$, if a high enough bias is applied between the leads,
the oscillator dynamics becomes non-trivial, with the possibility of switching between stationary
 states of different amplitudes. \cite{blanter}
At strong couplings, $E_p > \hbar \omega_0$, there is another kind of multistability that appears
at relatively small voltages, $eV < E_p$ -- the system can switch between the states corresponding
to (approximately) 1 and 0 electrons on the molecule.
This multistability and switching can be
described within the generalization of the Born-Oppenhemier approach to open systems.\cite{MHM}
In the metallic case the appearance of the multistability and the current suppression as a function of
the bias voltage is associated with a discontinuity of the current (when
the cotunnelling is neglected).\cite{fabio}

The slow (or ``classical") phonon strong coupling case is attractive since besides
switching between the different charge-ion states, it allows a read-out of the state
 by means of {\em
cotunneling} transport through the molecule.
In cotunneling, the charge state of the molecule changes only virtually for a
period of time determined by the energy uncertainty principle.  This time can
be much shorter than the vibration period, and thus the ionic configuration and
the average charge occupancy need not change.  On the other hand, in sequential
tunneling, the tunneling events between the leads and the molecule are
energy-conserving, with the rates determined by the Fermi's Golden rule.
Typically, cotunneling currents are much smaller than sequential ones since
they are higher order in the tunneling matrix element. However, if the
sequential tunneling is strongly suppressed by the Franck-Condon physics, the
cotunneling, which needs not be affected by it, may dominate.  In the case
$\Gamma < \omega_0$ and strong electron-ion coupling the role of
cotunneling was recently studied in Ref.~\onlinecite{andreev}, where it was found
that while it does not destroy the Franck-Condon blockade, it can dramatically
affect the low-voltage current and current noise, as well as the vibrational
dynamics.

The purpose of this work is to provide a unified self-consistent description of
the sequential and cotunneling transport regimes in the case of a molecular
switch in the ``classical'' regime $\Gamma >  \omega_0$ and $eV>\hbar \omega_0$.
This regime allows for a
systematic non-perturbative treatment for an arbitrary electron-ion coupling
strength.\cite{MHM}  We determine the dynamics of the vibrational degree of
freedom, the average current and current noise through the device, and the
switching times as functions of transport and gate voltages.  We also analyze
the role of extrinsic dissipation.

\section{Model}
\label{Sec:model}

We consider the model for a molecular switch proposed in
Ref.~\onlinecite{MHM,galperin}.  The molecule is modeled as a single
electronic level $\hat{d}$ strongly interacting with a vibrational
mode, $x$. It is located between two leads, from which electrons can tunnel into the
electronic level. The interaction is provided by the force $\lambda$
(typically of electrostatic origin) acting on the molecule. The system
is described by the Hamiltonian
\beqa
H &=& (\epsilon_0 + \lambda x) \hat{d}^\dag \hat{d} + \frac{p^2}{2m} + \frac{m\omega_0^2
x^2}{2}\\
&&+ \sum_{k,\alpha}{\epsilon_{k\alpha} \hat{c}^\dag_{k\alpha}\hat{c}_{k\alpha}} +
\sum_{k,\alpha}{t_{\alpha} (\hat{c}^\dag_{k\alpha}\hat{d} +
  \hat{d}^\dag \hat{c}_{k\alpha})},
\eeqa
where $\alpha$ is the lead index ($L$ or $R$) and $\hat{c}$ and
$\hat{d}$ are the electron
annihilation operators for the leads and local orbital, respectively.  We
consider the model for spinless electrons for simplicity. (Inclusion of spin
along with onsite Coulomb blockade should lead to qualitatively similar
results.)  The vibrational mode is characterized by the ``bare" frequency
$\omega_0$ and the effective mass $m$.  The displacement and coordinates are
described by the canonically conjugate operators $x$ and $p$.  The coupling
between the electronic level and the mode is characterized by the ``polaron"
energy $E_p = \lambda^2/(2 m\omega_0^2)$ and the coupling to the leads by
tunnel rate $\Gamma_\alpha = \pi \nu_\alpha t_\alpha^2/\hbar$, where $\nu_\alpha$ is
the density of states in lead $\alpha$.  In Refs. \onlinecite{MHM,galperin} it has
been shown that for strong enough coupling, $E_p/\hbar \gg \Gamma_L + \Gamma_R$, the
system can exhibit bi-stability, with one state corresponding to empty resonant
level and non-displaced mode $x$, and the other to occupied level and the mode
displaced by the amount $\sim \lambda/(m\omega_0^2)$.  In the previous work,
Ref.~\onlinecite{MHM}, current and current noise were determined in the regime
of small transport voltage, $|eV| \ll E_p$ (where $eV=\mu_L-\mu_R$)
in the approximately ``symmetric" situation, $\epsilon_0 \approx E_p$.
In the present work, we generalize the
previous results for current and current noise as well as determine the
behavior of the switching rates between the metastable states for arbitrary
transport and gate voltages.

When electrons are driven out of equilibrium by an applied transport voltage,
the dynamics of the vibrational mode becomes very simple, even for strong coupling
between the mode and electrons.
That is because when the characteristic
timescale for electronic subsystem becomes shorter than oscillator frequency $\omega_0$,
electrons appear to the mode as a ``high-temperature," albeit position dependent and
strongly coupled bath.
Physically, for any position $x$, the electronic bath adjusts (almost!)
instantaneously, in a manner analogous to how electrons adjust to the instantaneous positions of
ions in isolated molecules, as described by the Born-Oppenheimer approximation.
Indeed, as in the standard Born-Oppenheimer approximation in equilibrium bulk solids,
one effect of the non-equilibrium fast electronic environment is the modification
of the effective potential experienced by the mode; however, what is more, the electronic
subsystem, by virtue of being open, also provides force noise ({\em fluctuations}) and the
dissipation to the mode.
Since the force acting on the mechanical mode is simply $-\lambda n$,
where $n=\hat{d}^\dag \hat{d}$ is the occupation of the electronic mode, in order to
obtain the average force and its fluctuation it is enough to calculate
the average of $n$ and it's fluctuation (charge noise) for a given
static position $x$.
When a weak time dependence of $x(t)$ is included one finds that
a correction to the average of $n$ appears that is linear in $dx/dt$.
This last term corresponds to the dissipation induced by the
retardation of the electronic degrees of freedom, that do not
respond immediately to a change of $x$ (first non-adiabatic correction).\cite{armour}
It can also be traced to the ``quantum" nature of the charge
noise, i.e. a slight asymmetry between the charge noise at positive and
negative frequencies.\cite{aguado,gardiner,clerk}
As a result, the dynamics of the mode $x$ becomes essentially classical,
described by the Langevin equation,\cite{MHM}
\beq
    m\ddot x + A(x) \dot x  + m \omega_0^2 x = F(x) + \xi(t),
    \label{langevin}
\eeq
where the position-dependent force $F$, damping $A$, and the intensity of the
white noise $D$, $\langle\xi(t)\xi(t') \rangle = D(x) \delta(t-t')$ are related
to the electronic Green functions on the Keldysh contour as
\beqa
F(x) &=& -\frac{\lambda\hbar }{2 \pi i}\int{d\omega G_{fr}(\omega,x)},\\
A(x) &=& \frac{\lambda^2\hbar}{2\pi} \int{d\omega G_{fr}(\omega,x)\partial_\omega
G_{rf}(\omega,x)},\\
D(x) &=&\frac{\lambda^2 \hbar}{2\pi}\int{d\omega G_{fr}(\omega,x)
  G_{rf}(\omega,x)} \ .
\eeqa
The zero temperature Green functions (for the forward-reverse Keldysh time path) are
\beqa
    G_{fr}(\omega,x)
    &=&
    2i\frac{\hbar \Gamma_L \Theta(\mu_L - \hbar \omega) + \hbar \Gamma_R
    \Theta(\mu_R - \hbar \omega)}{(\hbar\omega -\epsilon_0 -\lambda x)^2 +\hbar^2\Gamma^2},
    \\
    G_{rf}(\omega,x)
    &=&
    -2i\frac{\hbar \Gamma_L \Theta(\hbar \omega - \mu_L) + \hbar \Gamma_R
    \Theta(\hbar \omega - \mu_R)}{(\hbar \omega -\epsilon_0 -\lambda x)^2 +\hbar^2\Gamma^2}.
\eeqa
Here $\Gamma = \Gamma_L + \Gamma_R$. These expressions are valid also at finite
but low temperatures such that $ k_B T< \hbar \Gamma$.
[At higher temperature
the step functions $\Theta(\epsilon)$ have to be replaced by Fermi functions $n_F(-\epsilon/k_B T)$.]
Therefore, at low temperatures, we obtain,
\beqa
    F(x)
    &=&
    -\frac{\lambda}{\pi\Gamma}\left[ \Gamma_L\left(\tan^{-1}\frac{\mu_L -
    \epsilon - \lambda x}{\hbar \Gamma}+\frac{\pi}{2}\right)\right. \nonumber\\
    &&+ \left. \Gamma_R\left(\tan^{-1}\frac{\mu_R - \epsilon -
    \lambda x}{\hbar\Gamma}+\frac{\pi}{2}\right)\right] \ ;
    \label{defF}
    \\
    A(x)
    &=& \frac{\lambda^2\Gamma \hbar^3}{\pi}\left\{\frac{\Gamma_L}{[(\mu_L - \epsilon_0
    -
    \lambda x)^2 + \hbar^2 \Gamma^2]^2}\right.\nonumber\\
    && + \left.\frac{\Gamma_R}{[(\mu_R -
    \epsilon_0 - \lambda x)^2 + \hbar^2\Gamma^2]^2}\right\} \ ;
    \label{defA}
    \\
    D(x)
    &=& \frac{\lambda^2 \Gamma_L \Gamma_R}{\pi \Gamma^3}\left(\tan^{-1} z +
    \frac{z}{z^2 + 1}\right)^{\frac{\mu_L - \epsilon_0 - \lambda
    x}{\hbar\Gamma}}_{\frac{\mu_R - \epsilon_0 - \lambda x}{\hbar
    \Gamma}} \ .
    \label{defD}
\eeqa
Note that the expression for the force is just $F = -\lambda n(x)$, where
$n(x)$ is the occupancy of the $d$ level for a fixed displacement $x$.  The
expression for $D$ is given for $\mu_L > \mu_R$, otherwise, the $\mu_L$ and
$\mu_R$ have to be interchanged.

\section{Current and Noise from the Fokker-Planck description}
\label{Sec:FP}

From the Langevin Eq. \refe{langevin} one can derive a Fokker-Plank equation
for the probability $\Pc(x,p,t)$ that at a given time $t$ the displacement and
the momentum of the vibrational are $x$ and $p=m \dot x$,
\beq
\partial_t \Pc =  -{p\over m} \partial_x \Pc - F(x) \partial_p \Pc + {A(x)\over m}
\, \partial_p(p\Pc) + {D(x)\over 2}\, \partial^2_p \Pc
\,.
\label{FokkerPlanck}
\eeq
This Fokker-Plank equation can be used to study both the stationary properties
of the system, as well as the time evolution from a given initial condition.

\subsection{Current}

Given our assumption about the separation between the slow ionic
-- vibrational -- and fast electronic -- tunneling -- timescales, the problem of
evaluating the stationary current reduced to the evaluation of the {\em
quasistationary} current averaged over the fast electronic times for a fixed
position $x$ and momentum $p$ of the mode, with the consequent averaging over
the stationary probability distribution, $\Pc(x,p)$.  In our case, the
quasistationary current through the molecule  depends then only on the position
$x$  (for $k_B T \ll \hbar \Gamma$),
\beq
    I(x)= {e \over 2 \pi} \int_{\mu_R}^{\mu_L} d\omega T(\omega,x) \ ,
\eeq
with
\beq
     T(\omega,x)={4 \Gamma_L \Gamma_R \over (\omega-\epsilon_o-\lambda x)^2+\Gamma^2}
     \,.
\eeq
The expectation value current is then simply
\beq
    I(t)=\int dx dp \Pc(x,p) I(x)
    \label{Iexp}
    \,.
\eeq
Solving the stationary \refE{FokkerPlanck} one can thus obtain the current
voltage characteristics for the device.

\subsection{Current noise}

We are also interested in the current noise:
\beq
    S(\omega) = \int dt e^{i \omega t} \qav{\tilde I(t) \tilde I(0)+ \tilde I(0) \tilde I(t)}
    \,,
    \label{Somega}
\eeq
where $\tilde I=\hat I- \qav{\hat I}$ and $\hat I$ is the current (quantum) operator.
Again, since in our problem we have a clear time-scale separation between the
vibrational and electronic degrees of freedom, we can distinguish two
contributions to the current noise.
The first is quasistationary (for a given position $x$) shot noise which arises
due to the discrete nature of the electron charge.
It has the usual form for a device with a single channel and
transparency $T(x,\omega)$\cite{lesovik},
\beq
    S_{shot}(\omega=0,x) = {2 e^2 \over \hbar} \int_{\mu_R}^{\mu_L} {d \omega \over 2 \pi}
    T(\omega,x) [1-T(\omega,x)]
    \,.
    \label{Sshot}
\eeq
The only change due to the presence of the oscillator is the fact that it must
be averaged over the position, in the same way as we have done for the average
current above.

\newcommand{\Lc}{{\cal L}}
\newcommand{\Ic}{{\cal I}}

The second more interesting type of noise is caused by the fluctuations of the
position $x$.  It occurs on a long time scale, and thus, at low frequencies, it
can be much more important than the standard electronic shot
noise.\cite{blanter}
When the typical electronic and mechanical fluctuation times are of the same
order of magnitude one has to take into account the correlation between the two
sources of fluctuations.\cite{pistolesiFCS}
However, for our system the
separation of the time scales makes these two noises additive and allows for
their separate evaluation without regard for one another.

To obtain the low frequency ``mechanical" contribution to the noise one needs
to consider the autocorrelator of the quasistationary current \refe{Iexp} at
different times.  This requires knowledge of the time-dependent solution of the
Fokker-Plank equation \refe{FokkerPlanck}.  The evolution of the probability
can be rewritten in a more compact form as
\beq
    \partial_t \Pc = \Lc \Pc
    \label{evolEq}
\eeq
where $\Lc$ is the Fokker-Planck operator, in this notation
$\Pc$ is a vector ($\Pc_i$) and $\Lc$ is a matrix ($\Lc_{ij}$).
The index $i=(x,p)$ represents all the stochastic variables in
discrete notations.
For instance, the current operator $\Ic$ is diagonal in the $i$
variables [cf. \refE{Iexp}] so that the average current can be written
simply as
\beq
    \qav{I} = \sum_i \Ic_i {v_0}_i = (w_0, \Ic v_0) \ .
\eeq
where $v_{ni}$ and $w_{ni}$ are the right- and left-eigenvectors of $\Lc$ with
eigenvalue $\lambda_n$ ($\Lc v_n  = \lambda_n v_n$ and $w_n^\dagger \Lc  =\lambda_n w_n^\dagger $).
If the eigenvalues are not degenerate then one can always choose the normalization
so that $(w_n,v_m)= \delta_{n,m}$.
The conservation of the probability implies that $\lambda_0=0$, and by
definition $v_0$ is the stationary solution and $w_{0i} = 1$.
The fluctuation operator for the current is
$
    \tilde \Ic = \Ic-\qav{\Ic}
$
in terms of which we can define the current fluctuations:
\beq
    S(t>0) \equiv \sum_{ij} \tilde\Ic_i U_{ij}(t) \tilde \Ic_j {v_0}_j
    .
\eeq
Here $U_{ij}(t)$ is the conditional evolution probability that the system evolves
from the state $j$ at time 0 to the state $j$ at time $t$.
It must satisfy the evolution equation \refe{evolEq} with the boundary condition $U_{ij}(0)=\delta_{ij}$.
By Laplace transform [$\hat U(s)=\int_0^{+\infty}U(t) e^{-s t} dt$
with $\mbox{Re}\ s>0$ and
$U(t)=\int_{a-i\infty}^{a+i\infty} (ds/2\pi i) \hat U(s) e^{s t}$, $a>0$]
we obtain
\beq
(s-\Lc)\hat U(s)=U(t=0)=1 \, .
\eeq
We can then calculate the noise spectrum by using the symmetry $S(t)=S(-t)$,
\beq
    S(\omega) =\hat S(s=-i\omega+0^+)+\hat S(s=i\omega+0^+)
\eeq
where $\hat S(s)$ is the Laplace transform of $S(t)$ and has the form
\beq
    \hat S(s) = \sum_{ij} \tilde \Ic_i (s-\Lc)^{-1}_{ij} \tilde \Ic_j v_{0j}
    \,.
\eeq
We thus obtain
\beq
    S(\omega) =
    -2\sum_{ij} \tilde \Ic_i \left({\Lc \over
      \omega^2+\Lc^2}\right)_{ij} \tilde \Ic_j v_{0j} \ .
    \label{SomegaM}
\eeq

\section{Relevant parameter range}

We assumed from the beginning that $\omega_0 \ll \Gamma$.  This ensures that
the electronic dynamics of the device is faster than the vibrational one.
The only remaining relevant energy scale is $E_p$, which we have to compare to
the other two parameters $\hbar \omega_0$ and $\hbar \Gamma$.
If $\Gamma \gg E_p/\hbar$ the switching effects are difficult to observe since
the boundaries of the Coulomb diamonds are blurred on a scale $ \hbar \Gamma$
much larger than the energy scale of the vibrational motion.
We thus will not investigate this limit, but shall concentrate on the opposite
one of $ \Gamma \ll E_p/\hbar$.

It is convenient at this point to rewrite the Fokker-Planck equation in
dimensionless form by introducing the variables $y=k x/\lambda$,
$\tau=t\omega_0$, $q=p k/\lambda\omega_0 m$.
\refE{FokkerPlanck} becomes
\beq
    \partial_\tau \Pc =  -q \partial_y \Pc - \Fc\, \partial_q \Pc+ \Ac
    \, \partial_q(q\Pc) + {\Dc\over 2}\, \partial^2_q \Pc
\eeq
with
\beqa
 \Fc(y)&=&-y-1/2 -{1\over \pi }\left[
\gamma_L \tan^{-1}\left({v_g+v/2-y\over\tilde\Gamma}\right)
    \right.
    \nonumber \\
    &&
    \left.+
    \gamma_R \tan^{-1}\left({v_g-v/2-y\over\tilde\Gamma}\right)
\right]
\eeqa
where $\gamma_i=\Gamma_i/\Gamma$.
\beqa
    \Ac(y)&=&{\tilde \omega {\tilde\Gamma}^2\over \pi}
    \left[{\gamma_L \over [(v_g+v/2-y)^2+{\tilde\Gamma}^2]^2} \right.
    \nonumber \\
    &&
    \left.
        +{\gamma_R \over [(v_g-v/2-y)^2+{\tilde\Gamma}^2]^2}
    \right]
\eeqa
\beq
 \Dc(y) = {\gamma_L \gamma_R\over \pi} {\tilde \omega \over \tilde \Gamma}
 \left(\tan^{-1}z+{z\over z^2+1}\right)_{v_g-v/2-y\over \tilde
 \Gamma}^{v_g+v/2-y \over \tilde \Gamma}
\eeq
We have also introduced the bias and gate voltages,
\beq
   \mu_L-\mu_R= 2 v E_p \,, \qquad (\mu_L+\mu_R)/2-\varepsilon= 2 v_g E_p
\eeq
and the dimensionless system parameters $\tilde \Gamma= 2\hbar \Gamma/E_p$ and
$\tilde\omega = 2\hbar \omega_0/E_p$.

We can now discuss the limit of interest $\omega_0 \ll \Gamma \ll
E_p/\hbar $.
The fluctuating and dissipative parts of the Fokker-Planck equation
(coefficients $\Ac$ and $\Dc$) are much smaller than the force term ($\Fc$)
since they are proportional to $\tilde \omega \ll 1$.
For $\tilde \omega \rightarrow 0$ the force term remains finite, while $\Ac$ and $\Dc$ vanish.
One therefore expects that the evolution of the system can be further
coarse-grained in time.  The system evolves under the influence of $\Fc$ most
of the time and thus conserves its effective energy defined by
$E_{eff}(y,q) = U_{eff}(y,q)+q^2$, with
\beq
    U_{eff}(y)=-\int^y dy' \Fc(y').
    \label{effPot}
\eeq
The effect of the small terms $\Ac$ and $\Dc$ is to produce a slow drift among
the nearby constant-energy orbits.
The stationary solution should then be a function of $E_{eff}(y,q)$ alone and
it is possible to reduce the Fokker-Planck equation to an energy
differential equation that in presence of a single minimum
has the analytical stationary solution
\beq
   \Qc(E) = {\cal N} e^{\int^E (\alpha(E')/\beta(E')) dE'}/\beta(E)
   \label{simpleQ}
   \,.
\eeq
Here ${\cal N}$ is a normalization factor and
$
 \Qc(E,\tau) = \int dy dq \delta(E-E_{eff}(y,q)) \Pc(y,q,\tau)
$. The coefficients $\alpha$ and $\beta$ are obtained by averaging a
combination of $\Ac$ and $\Dc$ on the trajectories of given constant effective
energy $E_{eff}(y,q)=E$ , as discussed in detail in Ref. \onlinecite{fabio}:
$
    \alpha=\qav{{\Dc(y)/2}-\Ac(y) q^2}_E$
and
$
    \beta = \qav{p^2 {\Dc(y)/2}}_E
$.
Note that in \refE{simpleQ} $\tilde \omega$ cancels out in the
exponential.
Thus the limit $\tilde \omega \rightarrow 0$ is well
defined for the stationary distribution of probability.
Obviously in this limit the time to reach the stationary state diverges
since it is linearly proportional to $\tilde \omega$.

When the potential can be approximated by a quadratic function around a local
minimum and the $y$ dependence of the coefficients $\Ac$ and $\Dc$ can be
neglected, the expression for the probability becomes
\beq
   \Qc(E) = {\cal N} e^{-E/T^*} \ ,
\eeq
where $T^*=2 \Dc(y_m)/\Ac(y_m)$ and $y_m$ is the position of the
local minimum.

\begin{figure}
\begin{center}
\includegraphics*[width=5cm,angle=-90]{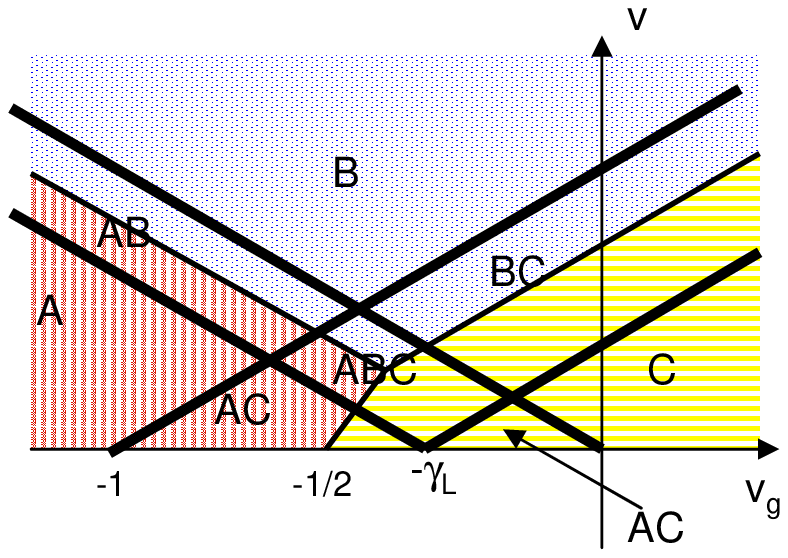}
\caption{
Regions in the $v-v_g$ plane of existence of the minima of $U_{eff}$ for $\tilde\Gamma \rightarrow 0$.
The letters $A$, $B$, and $C$ stand for the presence of a minimum at $y=-1$, $-\gamma_L$, and 0, respectively.
The plane is separated into three dashed regions according to which of the three extrema is the absolute minimum.
\label{diamonds}
}
\end{center}
\end{figure}

\begin{figure}
\centerline{
\includegraphics*[width=4.cm]{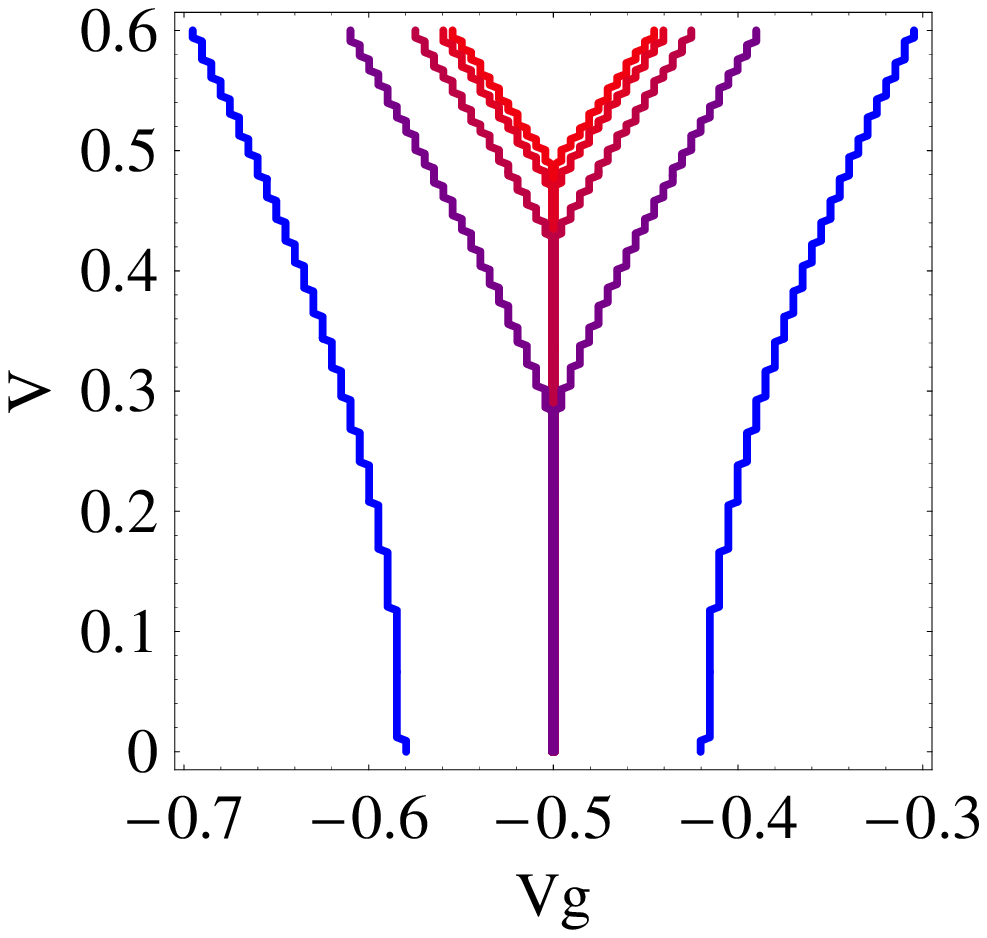}
\includegraphics*[width=4.cm]{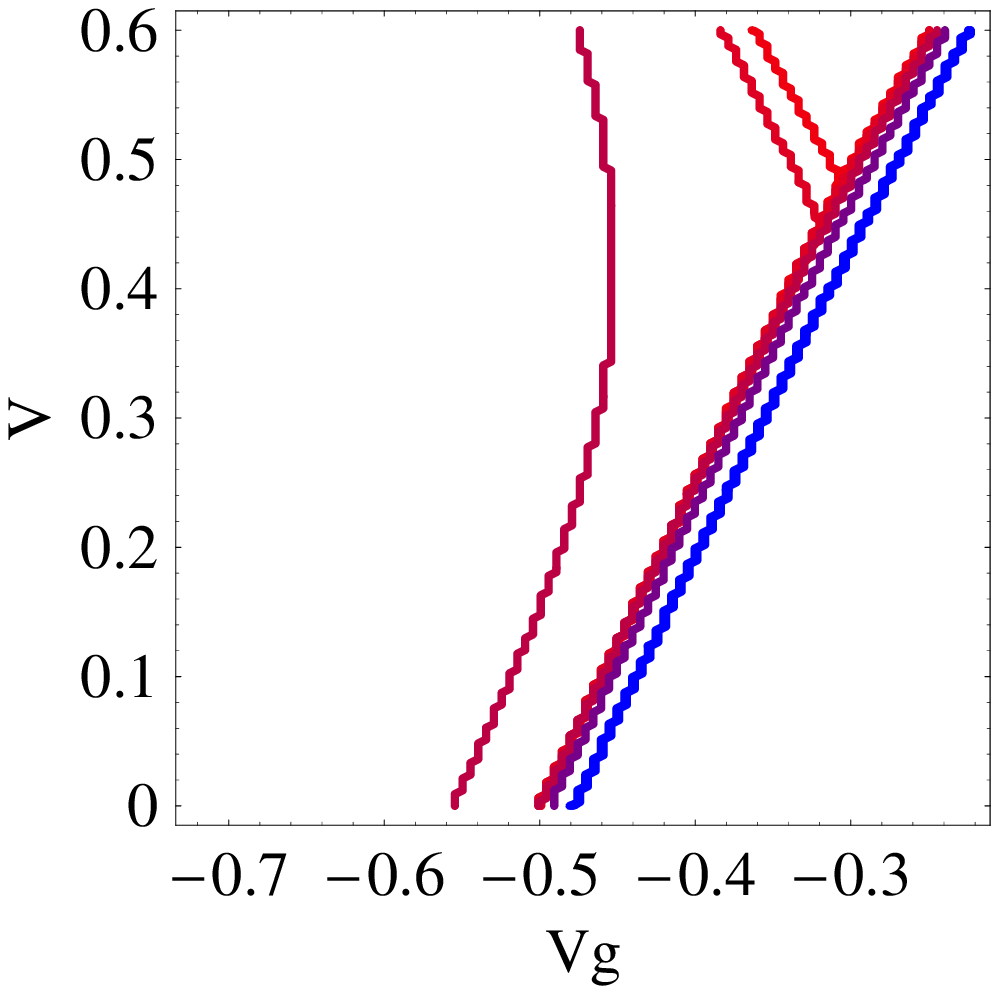}
}
\caption{
Regions of stability of the sequential tunneling solution for
$\gamma_L=1/2$ (left panel) and $\gamma_L=0.1$ (right panel),
and $\tilde \Gamma=0.02$, 0.04, 0.08, 0.16, and 0.30, (from
red to blue).
The region of sequential tunneling evolves from the small triangular shape
in the top of the plot (for $\tilde \Gamma$ small) to a large
trapezoidal shape (for large $\tilde \Gamma$) that touches the
$v=0$ axis.
The regions to the left and the right of the sequential tunneling are
``blocked" in the 0 or 1 occupation state, respectively.
\label{diamonds2} }
\end{figure}

Even if in the general case the stationary distribution is not determined
in such a simple way it is instructive to study the structure of $U_{eff}(y)$.
This is particularly simple for $\tilde\Gamma \ll 1$ since in this limit the
force becomes
\beq
 \Fc(y)=-y -\gamma_L \, \theta(v_g+v/2-y)-\gamma_R \theta(v_g-v/2-y)
 \,.
\eeq
It is then possible to show that the effective potential landscape
can show up to three minima at the positions $y=0$ for $v_g<-v/2$,
$y=-\gamma_L$ for $-v/2-\gamma_L<v_g<v/2-\gamma_L$, and
$y=0$ for $v_g>v/2-1$.
(For simplicity we consider only the $v>0$ case.)
The minimum at $y=-\gamma_L$ is due to the {\em sequential tunneling} for which
the average occupation of the dot is $0 \leq \gamma_L \leq 1$ (the
energy level lies in the bias window).
The other two minima correspond instead to classically-blocked
transport (thus {\em co-tunneling} is the dominant current mechanism),
either in the $n=0$ or $n=1$ state.
There are regions where two or three minima are present
at the same time.
One can show that for $-v/2-\gamma_L/2<v_g<v/2+1-\gamma_L/2$ and $v>1/2$ the
sequential tunneling minimum at $x=-\gamma_L$ is the absolute minimum.
In the rest of the plane either the blocked state 0, or the blocked state 1 are
true minima, the separation line between the two joins the point $v_g=-1/2$,
$v=0$ to the apex of the conducting region $v_g=-3/4+\gamma_R/2$ and $v=1/2$.
(cf. Figs. \ref{diamonds} and \ref{diamonds2}.)
%
%
%
For finite value of $\Gamma$ the stability diagram changes, the main difference
is the increase of the region of sequential tunneling that extends towards the
axis $v=0$, as shown in Fig. \ref{diamonds2}.
%
%

At low voltage and small $\tilde\Gamma$ one of
the two blocked states has the minimum energy.
For $\gamma_L=\gamma_R=1/2$ and $v_g=-1/2$
the effective temperature of these states
vanishes linearly with the bias voltage.
Thus for $v\rightarrow 0$ these are the ``cold" states.
The effective temperature at the sequential tunneling minimum ($x=-1/2$) is
$T^*=\pi v^4/2^4/\tilde\Gamma$, thus for small $\tilde \Gamma$ this state is
always ``hot."
Around $v=1/2$ the hot sequential tunneling state becomes the $U_{eff}$
minimum, and the system starts to fluctuate between the hot and cold states.
The dimensionless current $\tilde I=I/\Gamma e$ in the cold state is very small
$\sim \tilde \Gamma v$ while in the sequential tunneling regime it is
of the order one.
The fluctuations between these two states produces large telegraph current
noise, as discussed for small $v$ in Ref. \onlinecite{MHM}.

\begin{figure}
\centerline{
\includegraphics*[width=7cm]{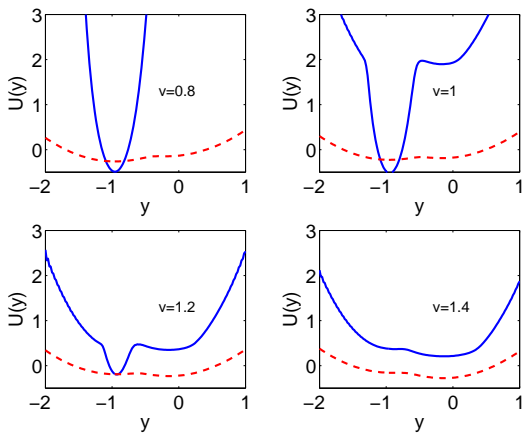}
} \caption{ Effective potential $U_{eff}(y)$ (red dashed) compared to $u_{eff}=-\ln
{\cal P}(y)$ (blue) for $\tilde\Gamma=0.08$, $\gamma_L=0.1$, $\tilde
\omega=10^{-3}$,$v_g=0$ and different values of $v$ as indicated in the panes.
The quantity $u_{eff}$ plays the role of an effective potential if $T^*$ was
constant.
Note in particular the case $v=1.2$ for which the absolute minimum of $U_{eff}$
is not the absolute minimum of $u_{eff}$ due to the fact that $T^*$ is much
lower in the other minimum. \label{Ueff} }
\end{figure}

The fact that the effective noise temperature varies as a function of the position can
lead to dramatic consequences.  In the conventional equilibrium statistical
mechanics, according to the Gibbs distribution, the lowest energy state is the
most probable one.  However, if the noise temperature varies as a function of
position, it may happen that the lowest energy state, if it experiences higher
temperature, may be less likely than a higher energy state that experiences
lower temperature.  We illustrate this point in Fig.~\ref{Ueff}, which compares
the naive effective potential profile $U_{eff}$ with the actual self-consistent
probability distribution.

We need to stress here, however, that we assume that the only environment that
is experienced by the mechanical mode so far is the non-equilibrium electronic
bath due to the attached leads.  If the dominant environment were extrinsic
(non-electronic), with a fixed temperature and the coupling strength, then the
effective potential would indeed uniquely determine the probabilities of
particular states.  We will come back to this point in Section \ref{extrinsic}.

%
%
%
%
In order to discuss the behavior of the device in the full range of parameters
here we resort to a numerical solution of the Fokker-Plank equation from which
we can determine both the current and the current noise of the device.
In the following section we discuss the numerical results.

\section{Numerical Results for the Current and zero-frequency Noise}
\label{Sect:IandS}

The expressions \refe{Iexp}, \refe{Sshot} and  \refe{SomegaM} can be used to
calculate the current and the noise of the device.  In general the analytical
evaluation of these expressions is not possible.   Numerically, the solutions
can be obtained by rewriting \refE{FokkerPlanck} on a discrete lattice
$(x,p)$ and replacing the derivatives with their finite differences
approximations.
If the equation is solved in a sufficiently large (e.g. rectangular) region in
the $x$-$p$ plane, one can use vanishing boundary conditions, since the
probability vanishes far from the origin.  The matrix corresponding to the
discretized Fokker-Plank operator $\Lc$ is very sparse and the numerical
solution is relatively easy for matrices of dimensions up to $10^5$.
The discretization step sizes $k \Delta x/\lambda$ and $\Delta p
k/\lambda\omega_0 m$ must be smaller than $\hbar \Gamma/E_P$ in order to have a good
convergence. This practically limits our numerical procedure to values of
$\hbar \Gamma/E_p > 0.01$.
%
%
%
%
%
%
\begin{figure}
\begin{center}
\includegraphics*[width=\larghezza]{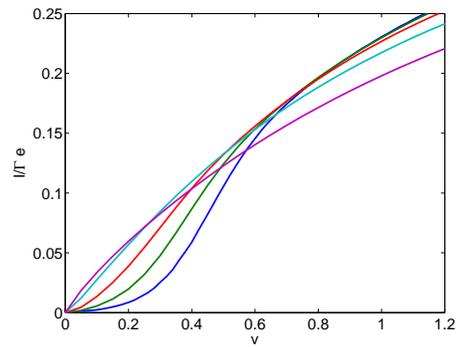}
\caption{
Current for $\tilde\Gamma=$ 0.02, 0.04, 0.08, 0.16, and 0.30,
from the lowest to the highest curve at low bias.
The other parameters are $\tilde \omega=10^{-3}$, $\gamma_L=1/2$, and $v_g=-0.5$.
\label{current2}
}
\end{center}
\end{figure}

We begin by considering the symmetric case, $\gamma_L=1/2$.
The current as a function of the voltage bias for different
values of $\tilde \Gamma$ is shown in Fig. \ref{current2}.
One can see that  for $\tilde \Gamma \rightarrow 0$ the current is
suppressed for $v<1/2$ and rises very rapidly for transport voltages exceeding
the threshold, as expected from the qualitative arguments given above.
Numerically is difficult to reduce $\tilde \Gamma$ further, but we expect that
for $\tilde \Gamma \rightarrow 0$ a discontinuity should appear as found in the
case when cotunnelling is negligible.\cite{fabio}
%

%
%
%
%
%
%
%
\begin{figure}
\begin{center}
\includegraphics*[width=\larghezza]{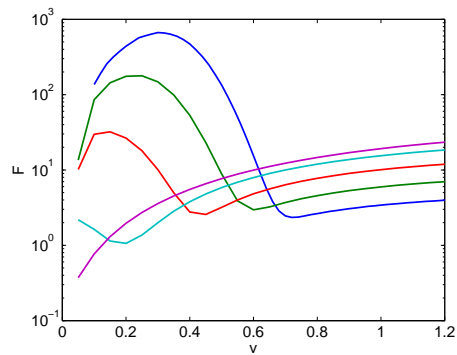}
\caption{
Fano factor of the current noise
in logarithmic scale
for $\tilde\Gamma=$ 0.02, 0.04, 0.08, 0.16, and 0.30, from
the lowest to the highest curve at large bias.
$\tilde \omega=10^{-3}$, $\gamma_L=1/2$ and $v_g=-0.5$.
\label{fano}
}
\end{center}
\end{figure}

In Fig. \ref{fano} we plot on a log scale the Fano factor [$F=S(\omega=0)/2eI$] of the
mechanically generated current noise (the standard shot noise contribution
is much smaller).
One can see that $F$ reaches huge values of the order of $10^3$, while it is
typically 1 for the purely electronic devices.
The maximum of the Fano factor appears slightly below the value of the voltage
where there is a crossover from the the cold to the hot minima; we will see
later that this corresponds to the value for which the switching rates between
the two minima are nearly the same.
Since the blocked minimum is colder than the sequential tunneling
minimum, this crossover happens before the hot minimum becomes
a true minimum.
Enhancement of noise in this device should serve as a strong
indication of the presence of mechanical oscillations.

%
%
%
%
%
%
\begin{figure}
\centerline{
\includegraphics*[width=7cm]{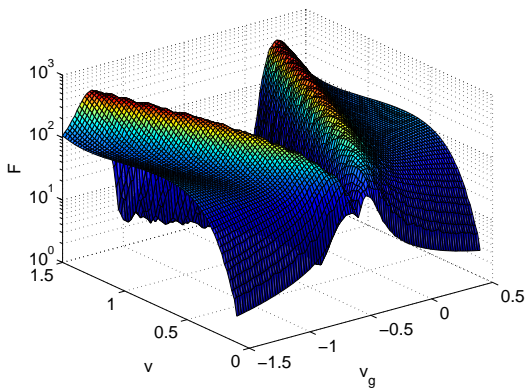}
}
\caption{{\bf Symmetric case.}
Fano factor for the induced current noise as a function of $v_g$ and $v$ for $\tilde\Gamma=0.08$, $\gamma_L=0.5$ and
$\tilde \omega=10^{-3}$.
\label{FanoDia}
}
\end{figure}
%

%
%
%
%
%
%
\begin{figure}
\centerline{
\includegraphics*[width=4cm]{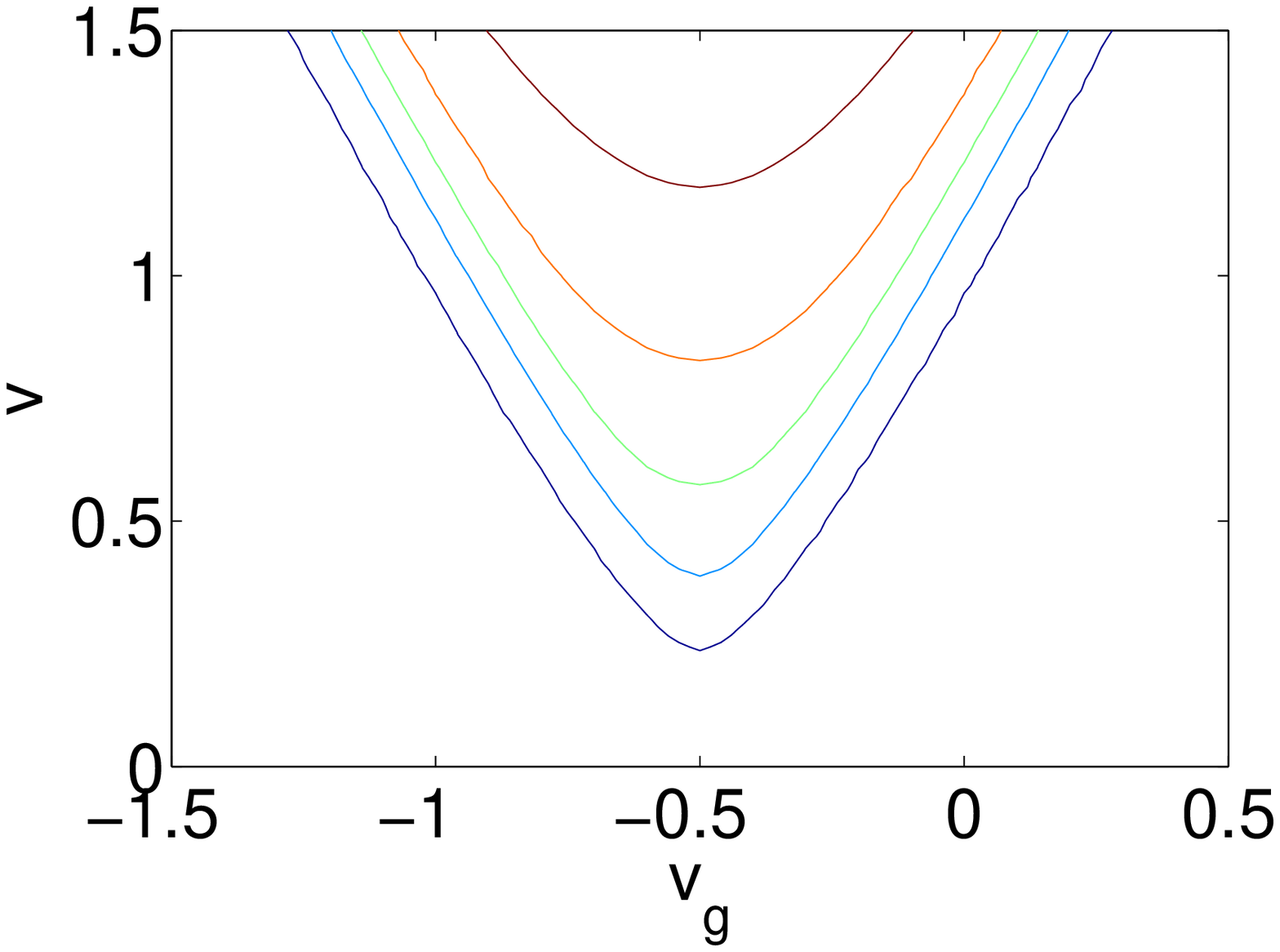}
\includegraphics*[width=4cm]{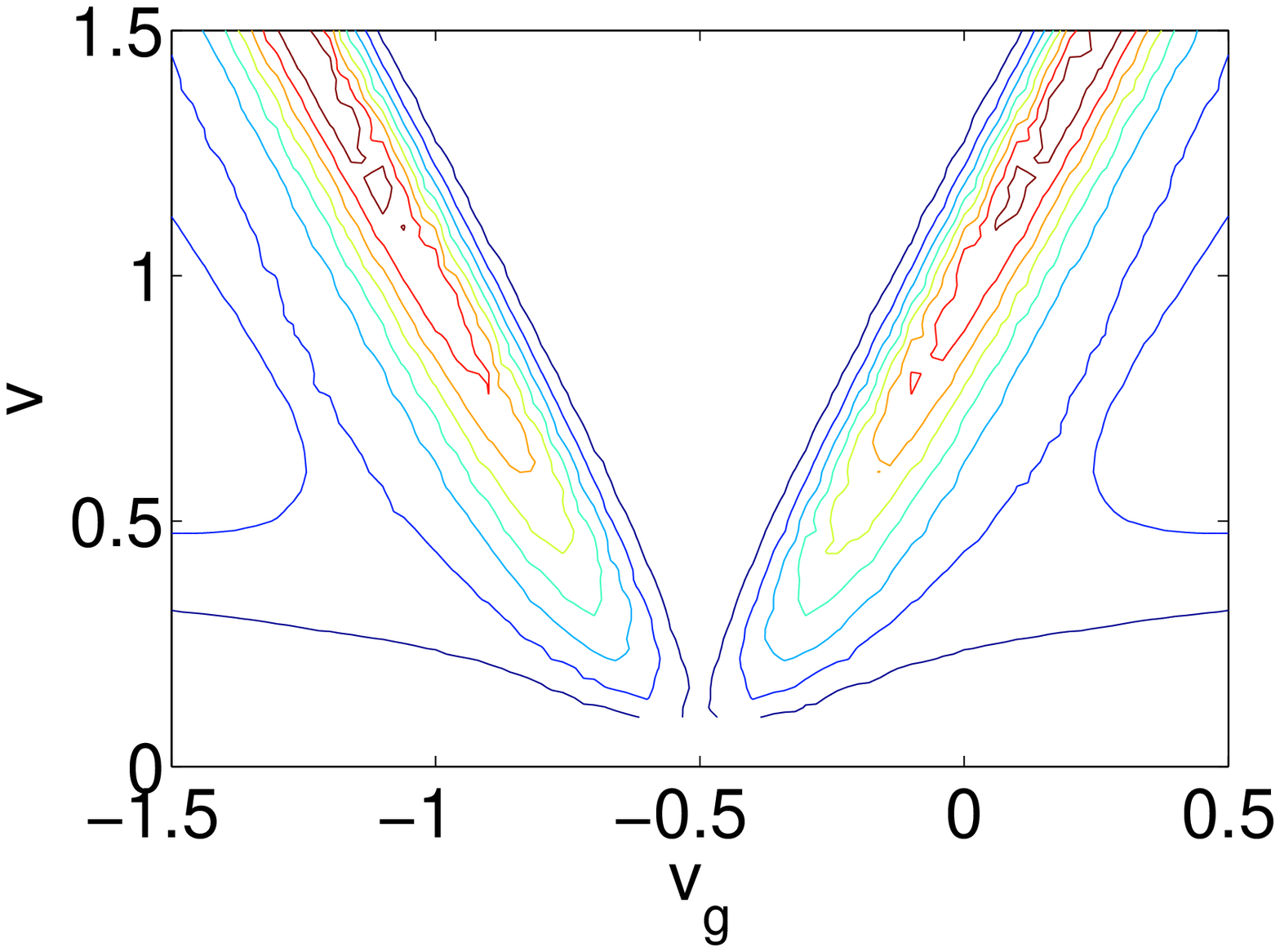}
}
\caption{
{\bf Symmetric case.}
Current and Fano factor for the mechanically induced current noise as a
function of $v_g$ and $v$ for $\tilde\Gamma=0.08$, $\gamma_L=0.5$ and
$\tilde \omega=10^{-3}$.
\label{SurfaceGamL05}
}
\end{figure}
%

%
%
%
%
%
%
\begin{figure}
\centerline{
\includegraphics*[width=7cm]{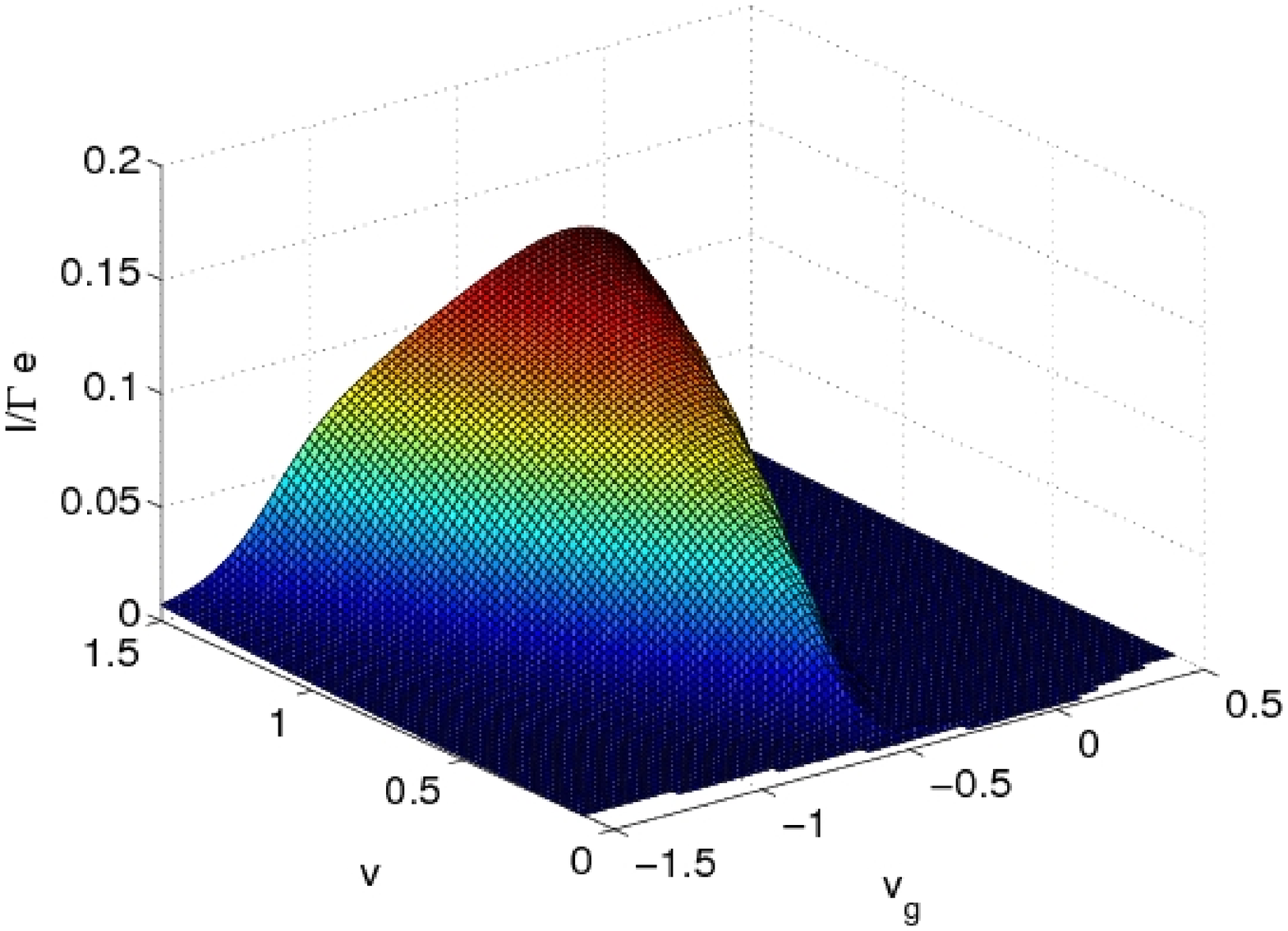}
}
\centerline{
\includegraphics*[width=7cm]{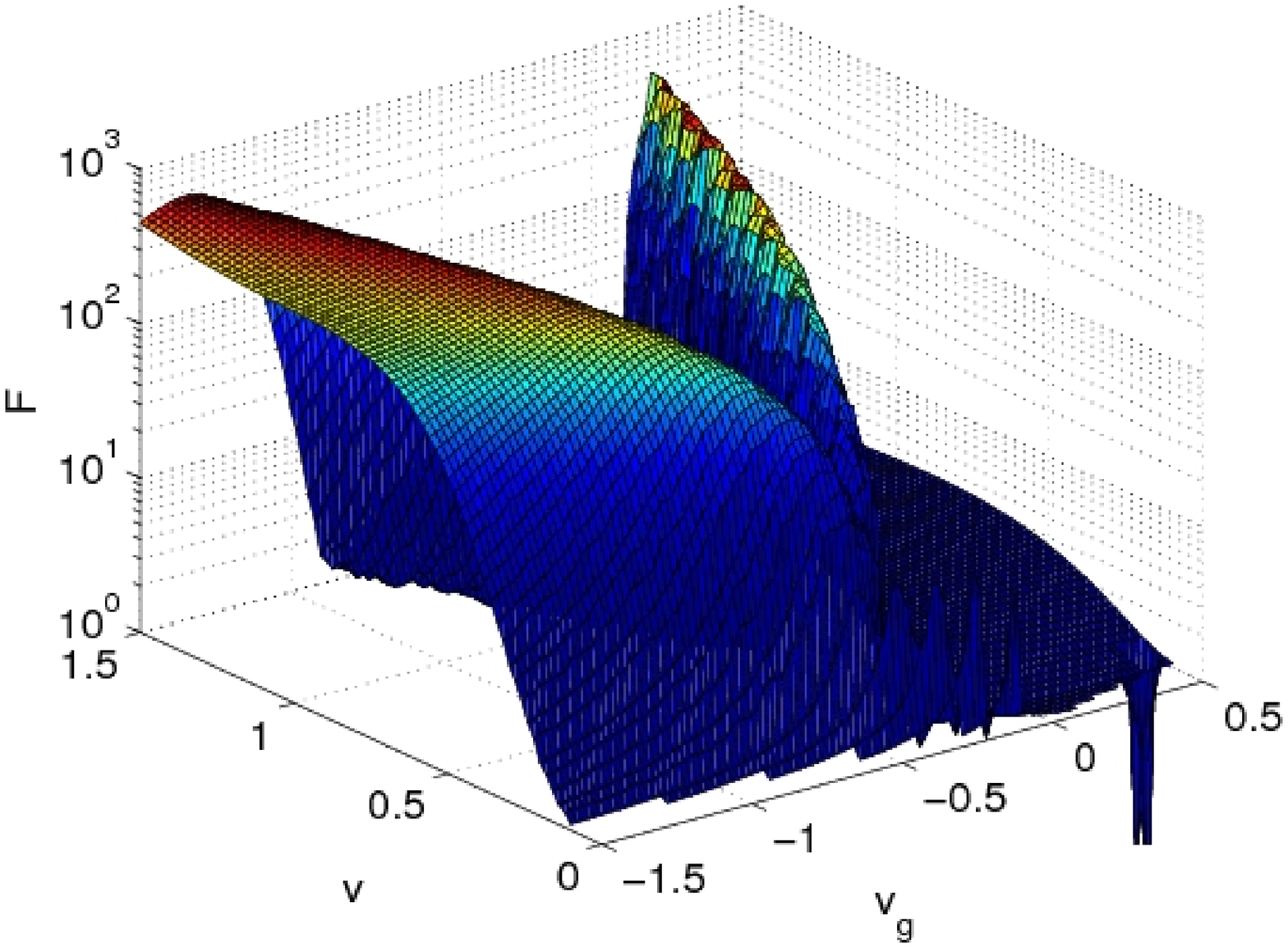}
}
\caption{{\bf Asymmetric case.}
Fano factor for the mechanically induced current noise as a function of $v_g$ and $v$ for $\tilde\Gamma=0.08$, $\gamma_L=0.1$ and
$\tilde \omega=10^{-3}$.
\label{SurfaceGamL01}
}
\end{figure}
%

%
%
%
%
%
%
\begin{figure}
\centerline{
\includegraphics*[width=4cm]{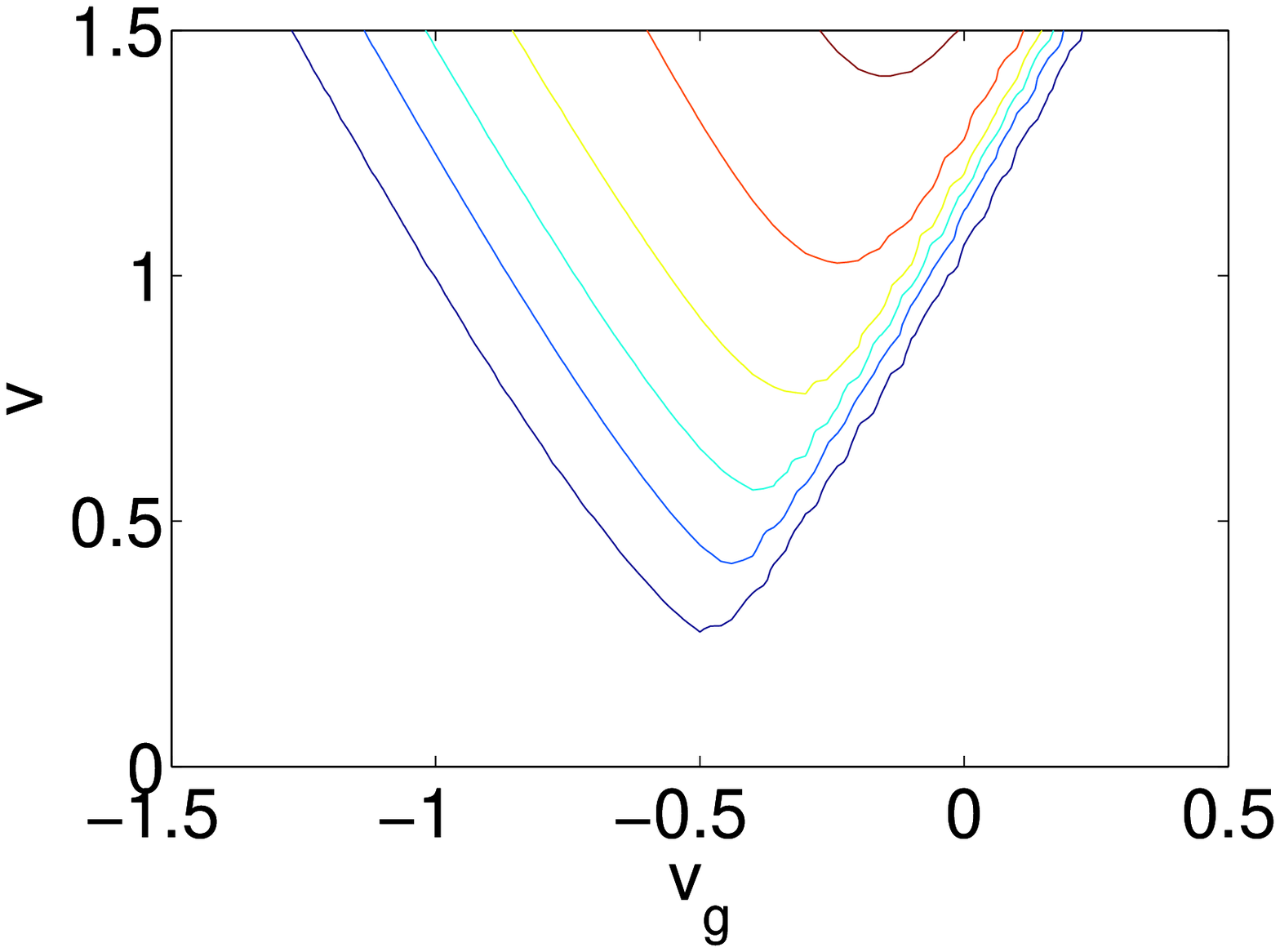}
\includegraphics*[width=4cm]{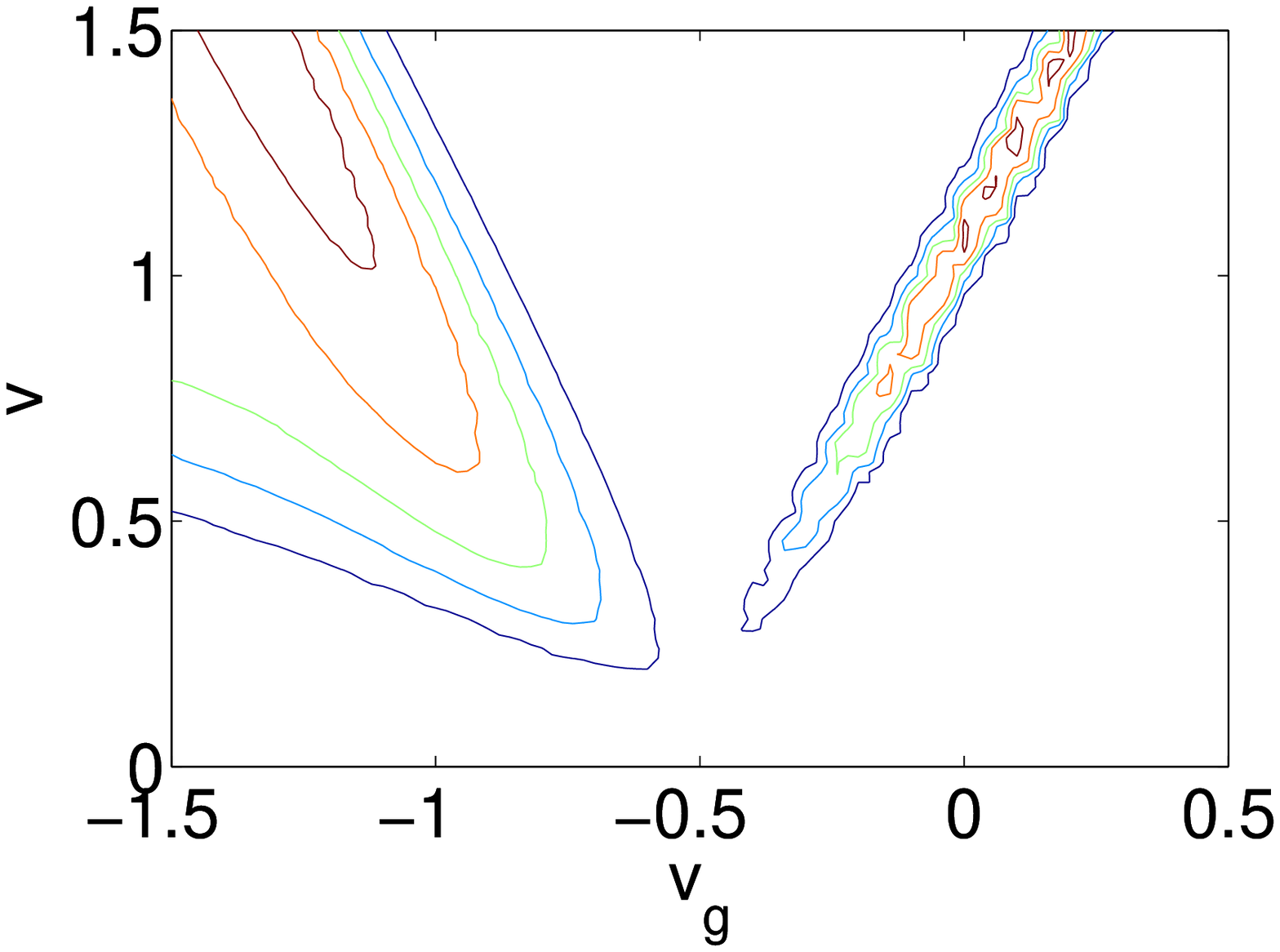}
}
\caption{
{\bf Asymmetric case.} Current and Fano factor for the mechanically induced current noise
as a function of $v_g$ and $v$ for $\tilde\Gamma=0.08$, $\gamma_L=0.1$ and
$\tilde \omega=10^{-3}$.
\label{FanoDia2}
}
\end{figure}
In Fig. \ref{FanoDia}, \ref{SurfaceGamL05}, \ref{SurfaceGamL01} and \ref{FanoDia2}
we show the behavior of the Fano factor in the plane $v_g-v$ for $\tilde \Gamma=0.08$
and $\gamma_L=0.5$ or 0.1.
Note that in the asymmetric case, Fig. \ref{SurfaceGamL01} and  \ref{FanoDia2},
there is a very sharp peak in the Fano factor if we increase the
bias voltage at fixed gate voltage greater than zero.
This structure appears at the threshold of the sequential tunnelling conducting
region.

\section{switching rate}
\label{Sect:switch}

In the previous sections we have studied the current and the current noise.
These quantities are the most readily accessible in transport measurements;
however, it is interesting also to investigate what is the typical switching
time $\tau_s$ between the two minima.
This quantity can give an indication if the telegraph noise could be detected
directly as a slow switching between discrete values of the average current.
For this to happen the switching time must be very long--at least comparable to
the average current measurement time (typically, in the experiment $\gtrsim 1\
\mu$s) .

To find a reliable estimate of $\tau_s$ we need to know the typical time
necessary for the system to jump from a local minimum of the effective
potential \refE{effPot} to a neighboring one.
This concept is well defined since the diffusion and damping term of the Fokker-Plack equation
are very small and the time evolution of the system on a short time scale
is controlled by the drift term.
Let us denote the value of the effective potential
at the local maximum separating the two minima of interest as $E_{max}$.
The region $\Omega$ on the $y$-$q$ plane around the minimum
defined by $E_{min} < E_{eff}(y,q) < E_{max}$ can be considered
as the trapping region.
If the system is at time 0 at the position $(y,q)$ inside
$\Omega$ we can estimate the average time to reach the boundary
of $\Omega$ ($\partial \Omega$) by solving the equation:
\beq
    \Lc^\dag \tau = -1
\eeq
with (absorbing) vanishing boundary conditions on $\partial\Omega$.\cite{BookFP}
Here $\tau$ stands for the function $\tau(y,q)$.
Since we are interested on the average time to leave the region we
average the escape time with the quasi-stationary distribution function.
The vanishing boundary conditions introduce a sink thus there is no
zero eigenvalue for the $\Lc$ operator with vanishing boundary conditions
on $\partial \Omega$.
We can nevertheless always identify the eigenvalue with the smallest real part and
call it $\lambda_0$: $\Lc v_0=\lambda_0 v_0$.
We thus obtain
\beq
    \qav{\tau} \equiv {(\tau,v_0) \over (1,v_0) }=
    -{1\over \lambda_0}
    \,.
\eeq
The inverse of the lowest eigenvalue gives the average switching time; this is
not surprising since the time evolution of the eigenstate $v_0$ is $e^{-t
\lambda_0}$.
It decays exponentially on a time scale $-1/\lambda_0$ due to the escape at the
boundaries of the region $\Omega$.

\begin{figure}
\centerline{
\includegraphics*[width=7cm]{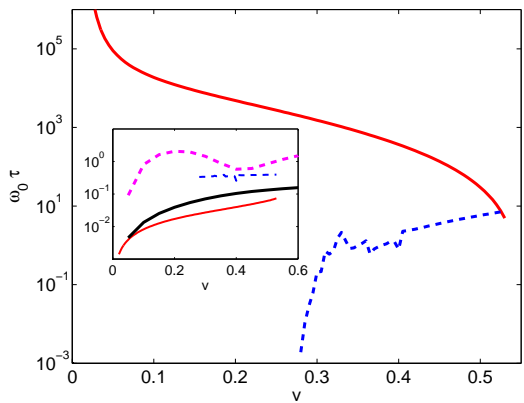}
}
\caption{{\bf Symmetric case.} $\gamma_L=1/2$, $\tilde\Gamma=0.08$, $v_g=-1/2$ and
$\tilde \omega=10^{-3}$.
Switching time between the two minima: red full line for the blocked
transport minima ($y=0$ or $y=-1$), and blue dashed line for the sequential
tunneling minimum.
In the inset: the current in each minimum (same notation of main plot) the average current
(black full line) and the current noise (magenta dashed line).
\label{Times}
}
\end{figure}

\begin{figure}
\centerline{
\includegraphics*[width=7cm]{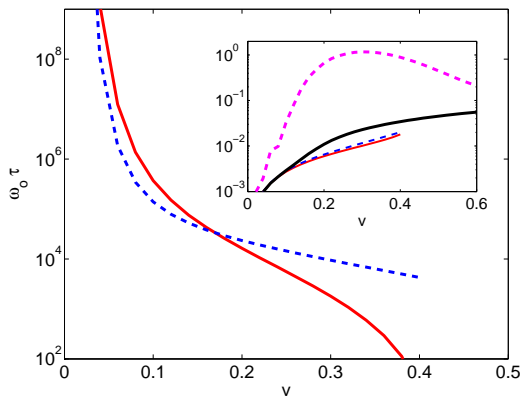}
}
\caption{{\bf Asymmetric case}. $\gamma_L=0.1$, $\tilde \Gamma=0.08$, $v_g=-1/2$, and
$\tilde \omega=10^{-3}$.
Same notations as as in Fig. \ref{Times}.
\label{Times2}
}
\end{figure}

We implemented numerically the calculation by solving the Fokker-Planck
equation in the energy-angle coordinates.
If $(y_o,0)$ is a minimum of the effective potential
with energy $E_{min}$, we rewrite the Fokker-Planck equation
in terms of the variables $E(x,q)=q^2/2+U_{eff}(x)$
and $\theta(x,q)=\arctan(q/(x-x_o))$.
In this way the boundary conditions
read  $\Pc(E=E_{max},\theta)=0$ for all values
of $\theta$.
The results are shown in Figs. \ref{Times} and \ref{Times2} for the
symmetric and asymmetric case, respectively.
%
%
%
%
%
%
%
%
%
%
%
%
%

Let us begin by discussing the symmetric case of Fig. \ref{Times}.
For small bias voltage only two minima are present, they are perfectly
symmetric and they correspond to two ``blocked" (classically-forbidden) current
state with $n=0$ or 1.
The switching time is very long, and the system switches between two blocked
states, each with very small cotunneling currents.  Since the cotunneling
currents for both minima in the symmetric state are the same, there is no
telegraph noise for small $v$.
As it can be seen from the value of result for the noise, the current
fluctuations are nevertheless high, and the reason is that to jump from one
minimum to the other the system has to pass through a series of states for
which current flows through the device is significant.
Moreover the slow fluctuations of the distribution function inside
each minimum are important for the noise as discussed in the following
Section \ref{Sect:Somega}.
The fact that the jumping times are so long may actually hinder the
observation of the jumps in a real experiment with finite measurement
time.
In a real device then the noise could be smaller in that case.
Increasing the voltage to $v\approx 0.28$, the sequential tunneling minimum at
$x=-.5$ appears and a true telegraph noise start to be present.
We see very clearly this in the escape times, which are no longer symmetric (we
plot the $y=-1$ and $y=-1/2$ minima escape times, the $y=0$ minimum has the same
behavior of the $y=-1$ minimum), and the average current at the minima also changes abruptly.
Even if the noise has a strong maximum near $v=0.28$ there is not
a dramatic increase {\em at} the appearance of the minimum.
The presence of the cotunneling smoothes the transition also for the noise that
has its maximum before the sequential minimum appears.
The switching time changes by 6 orders of magnitude in a very small range of
bias voltage.
Above $v\approx 0.53$ only the sequential tunneling minimum survives.

We consider now the asymmetric case of Fig. \ref{Times2}.
It is clear that the evolution of the escape times is very different
from the symmetric case.
In particular we consider the strongly asymmetric case of $\gamma_L=0.1$.  In
this case the sequential tunneling minimum merges with the blocked $n=0$
minimum, leading to a two minima landscape of the potential.
The consequence is that there is no abrupt appearance of a new minimum for some
value of the bias voltage; rather, the two minima are always present at the
same time till $v\approx 0.4$.
At low voltage the potential landscape is nearly symmetrical, both
minima are cold, but for the sequential tunneling one is characterized by
a slightly higher $T^*$ and thus its  escape time is shorter (dashed line
in Fig. \ref{Times2}).
Increasing the voltage, the height of the potential barrier for the blocked
state reduces, thus reducing the escape time.
At some point (in the case of Fig. \ref{Times2} for $v\approx 0.18$) the escape
time from the cold state become shorter than the escape time of the hot one,
since the the temperature has to be compared with the barrier, and at this
point the barrier height is smaller in the cold state.
Near the crossing region the noise shows a maximum, due to the
fact that the system spends nearly half of his time in each of the
two minima, with different average current.
Tuning $v$ one can thus cross from a region where the system is
trapped in one of the two minima, to a region where it jumps on a
relatively long time scale from one minimum to the other.
If the switching time scale becomes of the order of the response time of
the measuring apparatus it is in principle possible to observe directly
the fluctuation between the two values of the current.

\begin{figure}
\centerline{
\includegraphics*[width=7cm]{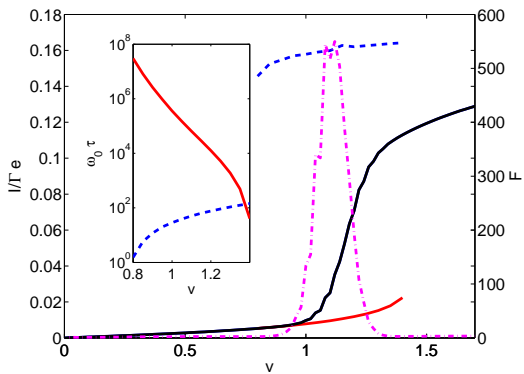}
}
\caption{
Average Current (black continuous line), Fano factor (magenta dot-dashed line), current in the two minima
(same notation as Fig. \ref{Times}) for
$\tilde \Gamma=0.08$, $\gamma_L=0.1$, $\tilde \omega = 10^{-3}$ at $v_g=0$.
In the inset the escape times from the two minima as a function of the bias voltage $v$.
}
\label{AllVg=0}
\end{figure}

This is even more pronounced if we follow the evolution of the current
at $v_g=0$. As can be seen in the contour plot of the Fano factors
(cfr. Fig. \ref{FanoDia2}),
in this way we will cross a very sharp peak of the Fano factor.
The results are shown in Fig. \ref{AllVg=0}. At low voltage only
a single nearly blocked state is present ($x=-1$ and $n=1$).
For $v\approx 0.8$ a new minimum appears at $x\approx -\gamma_L =-0.1$ that is
for the moment at higher energy and with a very small barrier.
The current associated to this minimum is much higher than the other,
and the system starts to switch between the two states.
The switching is very slow thus the noise is high.
Very rapidly as a function of $v$ the new local minimum becomes the
true minimum, and then the other minimum disappears.
%

%
%
%
%
%

\section{Frequency-dependence of the current noise}
\label{Sect:Somega}

\begin{figure}
\centerline{
\includegraphics*[width=7cm]{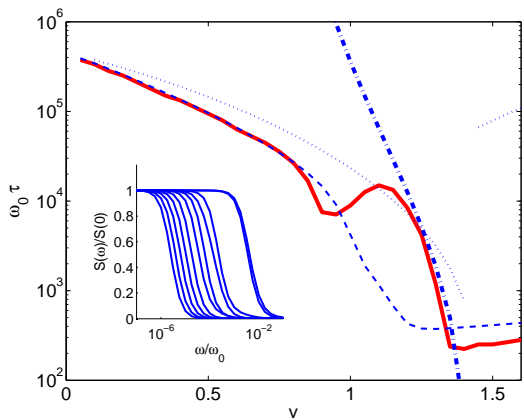}
}
\caption{Inset: Frequency dependence of the current noise for several values of the
bias voltage. From this data we estracted $\omega_c$ as the frequency at which
$S(\omega_c)=S(0)/2$.
Main plot: comparison of $\omega_0/\omega_c$ red full line, the escape time $\omega_0 \tau$
blue dot-dashed line, the friction coefficient $\omega_0m/A$ at the minimum dot line, and averaged
light dashed line.
The parameters are the same as Fig. \ref{AllVg=0}.
}
\label{Fig:Somega}
\end{figure}

Eq. (\ref{SomegaM}) derived above can be applied to study not only the zero
frequency noise, $S(\omega = 0)$, and the Fano factor, as we did in Section \ref{Sect:IandS},
 but also the current noise at an arbitrary frequency.
In this section we numerically
evaluate $S(\omega)$ and provide a qualitative explanation for the observed trends.
As we mentioned before, the shot noise contribution to the noise can be neglected
as far as the frequency considered is much smaller than $\Gamma$.
From the numerical calculations we find that the frequency dependence is characterized
by a single frequency scale, and approximately is Lorentzian peaked at $\omega = 0$.
This can be seen in the inset of Fig. \ref{Fig:Somega} where
we show $S(\omega)$ as a function of $\omega/\omega_0$
on a logarithmic scale for several values of the bias voltage $v$.
One can parameterize each curve by a single number, that we choose as
the frequency $\omega_c$ at which $S(\omega_c)=S(0)/2$.
It is instructive to compare the time scale $1/\omega_c$
with the energy dissipation and the switching timescales in various regimes.

At low voltages, since switching between the metastable minima is
exponentially slow, we anticipate that the low frequency ($\omega <
\omega_0$) current fluctuations will be determined by the energy
fluctuations within the single well in which the molecule spends most
of its time.
For a simple harmonic oscillator, the corresponding time scale is given by the inverse
damping coefficient.  For small energy fluctuations, the current changes with energy linearly.
Thus, current fluctuations will track the energy fluctuations, {\em i.e.} will be Lorentzian
with the width given by $A/m$.
To check this we plotted in Fig. \ref{Fig:Somega} the value
of $m\omega_0/A(x)$ evaluated at the minimum of the potential (dotted line).
There is a reasonable agreement for low voltage but, as expected, not for large voltages.
The reason is that at large $v$ the system becomes hot, and the energy dependence of $A$ cannot be neglected.
To address this issue, we calculated the average of $A(x)$
with the distribution function ${\cal P}(x)$ obtained by solving numerically
the stationary problem.
The result using thus obtained $A$ is shown as dashed line on the figure.
We find that it agrees very well with the $\omega_c$
extracted from the numerical calculation of $S(\omega)$, both at high and low voltages.
Note that at high voltage the energy dependence of $A$ is crucial to
understand the frequency response of the noise.
The effective temperature changes the average of $A$, and hence $\omega_c$ by
nearly three orders of magnitude.

In the intermediate transport voltage regime, $1 < v < 1.3$, the system switches between the two wells frequently.
 Therefore, we naturally expect that the timescale for the current noise should depend on the switching rate between
  the wells.  If each of the wells would corresponds to a fixed value of current the resulting noise would be a telegraph,
   with the Lorentzian lineshape and width given by the sum of the switching rates.  However, in each well as a function
   of energy current is not fixed.  In fact, the current increases gradually in the ``blocked" well as the energy the
    approaches the top of the barrier reaching the value $I \sim \Gamma$ near the top of the barrier.
On the other hand, in the well where transport is sequential, current remains approximately $I \sim \Gamma$ for any energy.
Therefore, one can naturally expect deviations from the simple telegraph behavior.  Indeed, we find that
the timescale $1/\omega_c$ tracks the escape time from the ``blocked" well (blue dot-dashed line in
 Fig. \ref{Fig:Somega}), which is the longer escape rate, and the fast escape from the ``hot" sequential
 well does not matter.  We therefore conclude that the noise is governed by the energy (and thus current)
  fluctuations within the cold (more probable) well, which also occur on the timescale comparable
   to the escape rate from it.

%
%
%
%
%
%

\section{role of extrinsic environmental dissipation}

\label{extrinsic}

As we discussed above most of the effects we found are
due to the non-equilibrium dynamics of the oscillator.
In order to improve our understanding of this fact, and to probe robustness of the
results to external perturbations we consider the influence of
extrinsic dissipation on the system.
This can be easily included in the model since the coupling
to an external bath implies only additional dissipation and fluctuation
on top of the intrinsic ones.
We assume that the system is damped due to the coupling to an external bath
at equilibrium at the temperature $T_{b}$.
The fluctuation and dissipation coming from this coupling satisfy the fluctuation dissipation theorem.
Thus the presence of the extrinsic damping induces the following
change in the variables $A$ and $D$ defined in Eqs. \refe{defA} and \refe{defD}:
$A\rightarrow A+\eta$ and $D\rightarrow D+ k_B T_b \eta/2$.
We present the numerical results for the dimensionless parameters
$\tilde \eta=\eta/m\omega_0$ and $\tilde T_b = 2k_B T/E_P$.
The numerical procedure remains unchanged.

%
%

%
\begin{figure}
\centerline{
\includegraphics*[width=7cm]{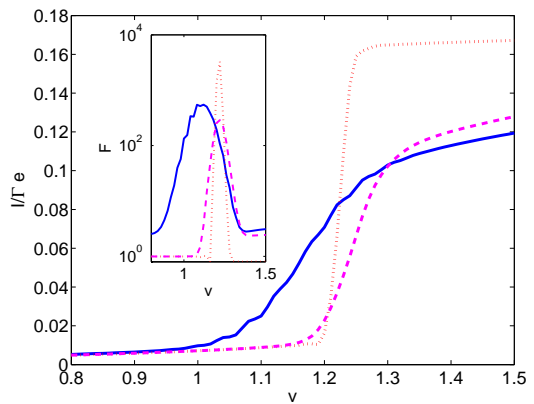}
}
\caption{
Voltage dependence of the current and Fano factor (inset) for different
values of the extrinsic dissipation: $\tilde \eta=0$, $10^{-3}$ and $0.1$.
The temperature of the external bath is $0.01$ in our dimensional
units, the other parameters are the same as Fig. \ref{AllVg=0}.
One can see that the current jump becomes sharper for stronger coupling
to the environment.
At the same time the Fano factor becomes sharper, thus a strong noise
region survives, but becomes very narrow when the external bath dominates.
}%
\label{CurrNoiseExtr}
\end{figure}
We show in Fig. \ref{CurrNoiseExtr} the behavior of the current and the noise
for the same parameters of Fig. \ref{AllVg=0} but at $\tilde T_b = 0.01$ and
for different values of the external dissipation.
The main feature that can be clearly seen is the sharpening of the
step for the current.
The external damping reduces the position fluctuations of the oscillator
thus reducing its ability to escape from the blocked regions of the parameters'
space.
On the other side if the oscillator is in a conducting region the
probability that it can fluctuate to regions of blocked transport is
smaller, thus the current is increased in the conducting regions and
reduced in the blocked regions, increasing the steepness of the step.
For the same reason the region of large noise is reduced.
We find that the value of the Fano factor remains actually very
large, but only in a very narrow range of bias voltages.
Increasing the coupling to the external bath reduces this windows
and thus finally rule out the possibility of observe it at all.

%
%
%
\begin{figure}
\centerline{
\includegraphics*[width=7cm]{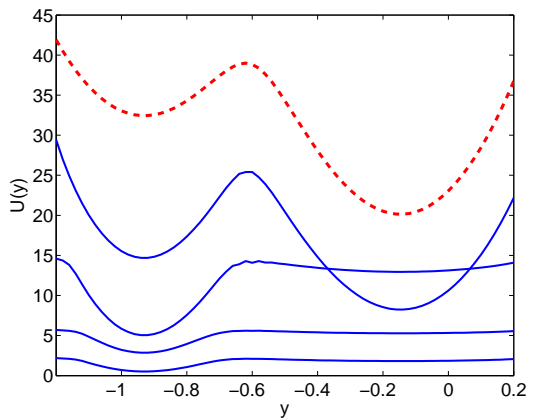}
}
\caption{
Effective potential $U_{eff}$ (red dashed) compared to
$-\ln({\cal P}(x))$ for different values of the extrinsic dissipation:
$\tilde \eta=10^{-5}$, $10^{-3.5}$, $10^{-2}$, and $10^{-0.5}$, from the
lowest to the highest curve.
The temperature of the external bath is $0.01$ in our dimensional
units, the other parameters are the same as Fig. \ref{AllVg=0} and
the curves are shifted and multiplied by a constant factor for clarity.
}
\label{ComparisonPot}
\end{figure}

A second interesting quantity to study is the distribution function ${\cal P}(x)$.
If the coupling to the environment dominates we expect that
${\cal P}(x) = {\rm const}\, e^{-U_{eff}(x)/\tilde T_b}$
to verify this fact we compare ${\cal U}(x)=-\ln {\cal P}(x)$ and
$U_{eff}(x)$ in Fig. \ref{ComparisonPot}.
We find that for small coupling first ${\cal U}(x)$ deviates even more
from the form of $U_{eff}$: the minimum in the cold regions deeps (left
minimum in the figure).
The reason is that the increase of the damping is more effective
in the cold region where both of damping and fluctuation are small.
In the hot region (right minimum in the figure)
the intrinsic fluctuation and dissipation is very large and for small
external damping there is no noticeable effect.
Increasing the coupling to the environment also the hot minimum
is cooled and the shape of $\cal U$ becomes similar to that of
$U_{eff}$ shown dashed in the plot.
This shows how relevant the non-equilibrium distribution of the position
is for the determination of the transport properties of the device.

\section{conclusions}

In this work our goal was to provide a unified description of the transport
properties of the strongly coupled non-equilibrium electron-ion system mimicking
a molecular device, in a broad range of parameters.  Our results are based on a
controlled theoretical approach, which only assumes that the vibrational frequency
 is the lowest energy scale in the problem.  In this regime, the vibrational mode
 experiences the effect of the electronic environment as a non-linear bath that has
 three interrelated manifestations: (1) Modification of the effective potential,
 including formation of up to two additional minima, (2) position-dependent force
 noise that drives the vibrational mode, and finally, (3) position-dependent dissipation.
We have self-consistently included the effect of tunneling electrons on the dynamics
of the vibrational mode, and the inverse effect of the vibrational mode on the electron
transport.  This enabled us to obtain the {\em average} transport characteristic of
the ``device," i.e. the dependence of the current on the transport and gate voltages,
as well as address the problem of current noise and mechanical switching between the
metastable states.  The agreement between the switching dynamics and the frequency
dependence of the current noise determined independently, enabled us to construct a
comprehensive but simple understanding of the combined electron-ion dynamics in
different transport regimes. In particular, the enhancement of current
noise may serve as an indicator of generation of mechanical motion,
and its magnitude and frequency dependence provide information on the
regime the molecular switching device is in and the values of relevant
parameters.


\section*{Acknowledgements}

We acknowledge useful discussions with A. Armour and M. Houzet.
This work has been supported by the French {\em Agence Nationale
de la Recherche} under contract ANR-06-JCJC-036 NEMESIS and Netherlands
Foundation for Fundamental Research on Matter (FOM).
The work at Los Alamos National Laboratory was carried out under the auspices
of the National Nuclear Security Administration of the
U.S. Department of Energy  under Contract No. DE-AC52-06NA25396 and
supported by the LANL/LDRD Program.

F. P. thanks A. Buzdin and his group for hospitality at the { \em Centre de Physique
Moleculaire Optique et Hertzienne} of Bordeaux (France) where part of this work has
been completed.

\end{document}